# Viscous Generalized Chaplygin Gas Interacting with $f(R,T)$ gravity.

E. H. Baffou$^{(a)*}$, I. G. Salako$^{(a,b)\dagger}$, M. J. S. Houndjo$^{(a,c)\ddagger}$

$^a$ *Institut de Mathématiques et de Sciences Physiques (IMSP),*
*01 BP 613, Porto-Novo, Bénin*
$^b$ *Département de Physique,*
*Université d'Agriculture de Kétou, BP 13 Kétou, Bénin*
$^c$ *Faculté des Sciences et Techniques de Natitingou - Université de Parakou - Bénin*

In this paper, we study in Friedmann-Robertson-Walker universe the interaction between the viscous generalized Chaplygin gas with $f(R,T)$ gravity, which is an arbitrary function of the Ricci scalar $R$ and the trace $T$ of the energy-momentum tensor. Assuming that the contents of universe is dominated by a generalized Chaplygin gas and dark energy, we obtained the modified Friedmann equations and also the time dependent energy density and pressure of dark energy due to the shear and bulk viscosities for three interacting models depending on an input parameter $Q$. Within the simple form of scale factor (power-law), we discuss the graphical representation of dark energy density parameter and investigate the shear and bulk viscosities effects on the accelerating expansion of the universe for each interacting model.

## I. INTRODUCTION

Recent observations of the supernovae type Ia (SNe Ia) [1], the cosmic microwave background radiation (CMBR) [2], the baryon acoustic oscillation (BAO) surveys[3], the large scale structure[4] and the weak lensing [5], clearly indicate that the universe is currently expanding with an accelerating rate. The invisible cosmic fluid called "dark energy" with a hugely negative pressure is responsible for this expansion. There are severals models to describe dark energy: cosmological constant [6], quintessence [7], phantom [8], quintom [9], tachyon[10], holographic dark energy [11], K-essence [12] and various models of Chaplygin gas.

Another proposal model for explaining the current acceleration of the universe which is confirmed by observational data [13]-[18] is to modify the gravitational theory. One of these theories is $f(R,T)$ theories of gravity proposed first by Harko et al.[40] and several work with important results have been found in such theory [19]-[25]. Furthermore, in the recently years, more and more cosmological observations suggest that our universe is permeated by imperfect fluid, in which the negative pressure, as was argued in [26]-[27], an effective pressure including bulk viscosity can play the role of an agent that drives the present acceleration of universe. The presence of viscosity in the fluid has been used to study dynamics of homogeneous cosmological models, and has application in studying the universe evolution. It is found that viscosity effects are viable at low redshifts, which observe negative pressure for the cosmic expansion with suitable viscosity coefficient. The CMBR observations indicate an isotropic universe, leading to bulk viscosity where the shear viscosity is neglected [28]. Long before the direct observational evidence through the SN Ia data, the indication of a viscosity dominated late epoch of accelerating expansion of the universe was already mentioned [29]. Many recently both bulk viscous effect and Chaplygin gas in FRW cosmology for the case of flat space considered and Friedmann equation due to Chaplygin gas which has bulk viscosity modified [30]. The viscous generalized Chaplygin gas is widely studied model among those proposed to describe the observed accelerated expansion of the universe. The effects of viscous fluid in modified gravity theories are analyzed to display accelerating expansion [31]-[33]. In recent years, there has been lot of interest results of the viscous fluid in modified gravity theories: Johri and Sudarshan [34] pointed out that the presence of bulk viscosity leads to inflationary universe in Brans-Dicke theory of gravitation. Das and Ali [35] studied axially symmetric Bianchi type $I$ bulk viscous cosmological model with time varying gravitational and cosmological constant. Sharif and Shamaila Rani [36] studied the bulk viscosity taking dust matter in the generalized teleparallel gravity. Naidu et al [37] studied Bianchi type-V cosmological model in $f(R,T)$ gravity when the source for energy momentum tensor is a bulk viscous fluid containing one dimensional cosmic strings. Kiran and Reddy [38] obtained that Bianchi type-III bulk viscous string cosmological model does not exist in $f(R,T)$ gravity and degenerates into vacuum model of general relativity. A viscous cosmology with matter creation in modified $f(R,T)$

---
* e-mail:baffouhet@gmail.com
† e-mail:inessalako@gmail.com
‡ e-mail: sthoundjo@yahoo.fr

gravity was discussed by Pankaj and Singh [39].

Inspired by the above discussion and investigations in modified theories of gravity, the idea is to studying in this framework the generalized Chaplygin gas Interacting with $f(R,T)$ gravity in presence of shear and bulk viscosities fluids. The paper is organized as follows: Sec. 2 provides basic formalism and discussion about the field equations of $f(R,T)$ gravity. In next section, we introduce briefly viscous cosmology in $f(R,T)$. In Sec. 4, we consider interaction of generalized Chaplygin gas with $f(R,T)$ gravity in presence of shear and bulk viscosities fluids and study effect of viscosity on the cosmological parameters namely the density energy of dark energy for three interacting model in power law of scale factor. The last section summarizes the results.

## II. FIELD EQUATIONS IN $f(R,T)$ GRAVITY

In the $f(R,T)$ background, the action is defined as [40]

$$S = \int \sqrt{-g} d^4x \left[ \frac{1}{2\kappa^2} f(R,T) + \mathcal{L}_m \right],  \tag{1}$$

where $f(R,T)$ is the arbitrary function on the curvature scalar $R$ and the trace $T$ of the energy-momentum tensor, $\mathcal{L}_m$ the density Lagrangian of the matter contents, and $\kappa^2 = 8\pi\mathcal{G}$, $\mathcal{G}$ being the gravitation constant.

The energy-momentum tensor of the matter is given by

$$T_{\mu\nu} = -\frac{2}{\sqrt{-g}} \frac{\delta(\sqrt{-g}\mathcal{L}_m)}{\delta g^{\mu\nu}}. \tag{2}$$

We assume that the matter Lagrangian density depends only the components of the metric tensor, and not its derivatives, so that on gets

$$T_{\mu\nu} = g_{\mu\nu} \mathcal{L}_m - \frac{2\partial \mathcal{L}_m}{\partial g^{\mu\nu}}. \tag{3}$$

By varying the action (1) with respect the components metric, we obtain the field equations in the metric $f(R,T)$ formalism defined as

$$f_R R_{\mu\nu} - \frac{1}{2} g_{\mu\nu} f(R,T) + (g_{\mu\nu} \Box - \nabla_\mu \nabla_\nu) f_R = \kappa^2 T_{\mu\nu} - f_T(T_{\mu\nu} + \Theta_{\mu\nu}), \tag{4}$$

where the tensor $\Theta_{\mu\nu}$ is evaluated as

$$\Theta_{\mu\nu} \equiv g^{\alpha\beta} \frac{\delta T_{\alpha\beta}}{\delta g^{\mu\nu}} = -2T_{\mu\nu} + g_{\mu\nu} \mathcal{L}_m - 2g^{\alpha\beta} \frac{\partial^2 \mathcal{L}_m}{\partial g^{\mu\nu} \partial g^{\alpha\beta}}. \tag{5}$$

In the expression (4), $f_R = \frac{\partial f(R,T)}{\partial R}$, $f_T = \frac{\partial f(R,T)}{\partial T}$ and the d'Alembert operator $\Box = g^{\mu\nu} \nabla_\mu \nabla_\nu$. It is important to note that the field equations in $f(R,T)$ gravity also depend on the physical nature of the matter field through the tensor $\Theta_{\mu\nu}$. Hence in the case of $f(R,T)$ gravity depending on the nature of the matter source, we obtain several theoretical models corresponding to each choice of $f(R,T)$. We chosen in this work the simplity form $f(R,T) = R + 2f(T)$ [40] and then, the field equations (4) becomes

$$R_{\mu\nu} - \frac{1}{2} R g_{\mu\nu} = \kappa^2 T_{\mu\nu} - 2f_T(T_{\mu\nu} + \Theta_{\mu\nu}) + g_{\mu\nu} f(T). \tag{6}$$

## III. VISCOUS COSMOLOGY IN $f(R,T)$ GRAVITY

We consider in this paper the matter content as generalized Chaplygin gas (assumed as dark matter) which has shear and bulk viscosities and we neglected the contribution of other components. We know that at the current time, the universe is homogeneous, isotropic at large scale and described by the flat Friedmann-Robertson-Walker (FRW) metric

$$ds^2 = dt^2 - a(t)^2 \left[ dx^2 + dy^2 + dz^2 \right], \tag{7}$$

where $a(t)$ represents the scale factor. Therefore, the energy-momentum tensor corresponding to the shear and bulk viscous fluid and generalized Chaplygin gas can be written as [41–47]

$$T_{\mu\nu} = (\rho + \bar{p}) u_\mu u_\nu - \bar{p} g_{\mu\nu}, \tag{8}$$

where $\rho$ and $\bar{p}$ are the energy density and the pressure of the matter and $u_\mu$ is the four-velocity vector with normalization condition $u_\mu u^\mu = 1$.
According the assumption, the pressure $\bar{p}$ of the matter is decomposed as

$$\bar{p} = p - 3\xi H + 2\eta H, \tag{9}$$

where $\xi$ and $\eta$ are bulk and shear viscous coefficients, respectively; $p$ being the pressure of the generalized Chaplygin gas (GCG) assumed as dark matter.

The Lagrangian density may be chosen as $\mathcal{L}_m = -\bar{p}$ and Eq. (5) reduces to

$$\Theta_{\mu\nu} = -2T_{\mu\nu} - g_{\mu\nu}\bar{p}. \tag{10}$$

Hence, the field Eq.(6) when the matter is considered as generalized Chaplygin gas which are shear and bulk viscosities can be written as

$$R_{\mu\nu} - \frac{1}{2} R g_{\mu\nu} = \kappa^2 T_{\mu\nu} + g_{\mu\nu} f(T) + 2 f_T (T_{\mu\nu} + g_{\mu\nu} \bar{p}). \tag{11}$$

By employing interacting model, we considered the total energy density and the total pressure of universe as the combination of components of a generalized Chaplygin gas and dark energy. Therefore, within equations (7) and (8), the corresponding modified Friedmann equations yields

$$3H^2 = \rho + f(T) + 2f_T (\rho + \bar{p}) = \rho_{total}, \tag{12}$$

$$-2\dot{H} - 3H^2 = \bar{p} - f(T) = p_{total}, \tag{13}$$

where we setted $\kappa^2 = 1$.

## IV. INTERACTING $f(R,T)$ GRAVITY WITH VISCOUS GENERALIZED CHAPLYGIN GAS

One of the recent cosmological models which is based on the use of exotic type of perfect fluid suggests that our universe filled with the generalized Chaplygin gas (GCG) with the following equation of state,

$$p = -\frac{B}{\rho^\alpha}, \tag{14}$$

where $B$ is a positive constant and $0 < \alpha \leq 1$; $\alpha = 1$ gives pure Chaplygin gas equation of state. Then, by considering the contents of the universe as generalized Chaplygin gas and dark energy components, the energy density and the pressure of all fluids can be written as

$$\rho_{tot} = \rho_{DE} + \rho, \tag{15}$$

$$p_{tot} = p_{DE} + p. \tag{16}$$

The continuity equation of conservation for each dark components fluids takes the following form

$$\dot{\rho}_{DE} + 3H(\rho_{DE} + p_{DE}) = -Q, \tag{17}$$

$$\dot{\rho} + 3H(\rho + p) = Q, \tag{18}$$

where the interaction term $Q$ corresponds to energy between dark energy and the generalized Chaplygin gas, which usually can be considered as $Q = 3Hb\rho$, $Q = 3Hb\rho_{DE}$, or $Q = 3Hb\rho_{tot}$; $b$ being the coupling constant. We have other interaction term which can be expressed in time derivatives of energy density, such as $Q = \mu\dot{\rho}$, $Q = \mu\dot{\rho}_{DE}$, or $Q = \mu \dot{\rho}_{tot}$. These type of interactions are either positive or negative and cannot change sign. However, a sign-changeable interaction of the following form is more interesting [48–52]

$$Q = q(\mu \dot{\rho} + 3bH\rho), \tag{19}$$

where $\mu$ and $b$ are constants and $q$ is the deceleration parameter expressed as

$$q = -1 - \frac{\dot{H}}{H^2}. \tag{20}$$

Note that when the universe changes from deceleration phase $q > 0$ to acceleration phase $q < 0$, this type of interaction can change its sign. According to the accelerating expansion of the universe, we choose in the present work three interacting form of model $Q$ to describe the evolution of the cosmological parameters in the universe dominated by viscous generalized Chaplygin gas and dark energy.

### A. Interacting model $Q = 3Hb\rho$

However, by substituting the interacting model $Q$ and Eq.(14) into Eq.(18), one gets in the absence of shear and bulk viscosities coefficients $\xi$ and $\eta$, the energy density of generalized Chaplygin gas as

$$\rho = \left[-\frac{B}{\gamma} + \frac{C_0}{\gamma} a^{3\gamma(1+\alpha)} e^{\gamma(1+\alpha)}\right]^{\frac{1}{1+\alpha}}, \tag{21}$$

where $\gamma = b - 1$ and $C_0$ is an integration constant. Furthermore, we deduct from Eq.(14), the pressure of generalized Chaplygin gas as

$$p = -B\left[-\frac{B}{\gamma} + \frac{C_0}{\gamma} a^{3\gamma(1+\alpha)} e^{\gamma(1+\alpha)}\right]^{-\frac{\alpha}{1+\alpha}}. \tag{22}$$

Also, the case of $\alpha = 1$ corresponding to pure Chaplygin gas yields,

$$\rho = \left[-\frac{B}{\gamma} + \frac{C_0}{\gamma} a^{6\gamma} e^{2\gamma}\right]^{\frac{1}{2}}. \tag{23}$$

$$p = -B\left[-\frac{B}{\gamma} + \frac{C_0}{\gamma} a^{6\gamma} e^{2\gamma}\right]^{-\frac{1}{2}}. \tag{24}$$

Within the modified Friedmann equations (12) and (13), the energy density and the pressure of dark energy takes respectively the following form

$$\rho_{DE} = f(T) + 2f_T(\rho + \bar{p}), \tag{25}$$

$$p_{DE} = -f(T) - 3\xi H + 2\eta H. \tag{26}$$

Moreover, the equation of state parameter $w_{DE}$ of dark energy is defined as

$$\begin{aligned} w_{DE} &= \frac{p_{DE}}{\rho_{DE}} \\ &= \frac{-f(T) - 3\xi H + 2\eta H}{f(T) + 2f_T(\rho + \bar{p})}. \end{aligned} \tag{27}$$

By assuming that the model $f(T)$ can be chosen as

$$f(T) = T^\beta, \tag{28}$$

where $T$ is the trace of energy-momentum tensor (8) and $\beta$ is constant, equations (25) and (26) becomes, respectively

$$\rho_{DE} = \left[\rho - 3p + 9\xi H - 6\eta H\right]^{\beta-1} \left(\rho(1 + 2\beta) + (2\beta + 3)(3\xi H - 2\eta H - p)\right), \tag{29}$$

$$p_{DE} = -\left[\rho - 3p + 9\xi H - 6\eta H\right]^{\beta} - 3\xi H + 2\eta H. \tag{30}$$

We consider in next section one special case of the scale factor (power law) in terms of cosmic time as follows

$$a \propto t^n, \tag{31}$$

to describe the evolution of cosmological parameters. It is possible to consider another models of given scale factor such as emergent, intermediate and logamediate scenarios, $a(t) = a_0(A + e^{kt})^m$, $a(t) = e^{\lambda t^{\zeta}}$ and $a(t) = e^{x(\ln(t))^{\delta}}$, respectively, which are based on different eras in the evolutionary process of the universe. To do so, we interest at evolution of the dark sector ie the energy density $\rho_{DE}$ and the pressure $p_{DE}$ of dark energy. Substituting the equations (21), (22) and (31) in (29) and (30), we obtain the following expressions

$$\rho_{DE} = \left(\left[-\frac{B}{\gamma} + \frac{C_0}{\gamma}t^{3n\gamma(1+\alpha)}e^{\gamma(1+\alpha)}\right]^{\frac{1}{1+\alpha}} + \frac{3B}{\left[-\frac{B}{\gamma} + \frac{C_0}{\gamma}t^{3n\gamma(1+\alpha)}e^{\gamma(1+\alpha)}\right]^{\frac{\alpha}{1+\alpha}}} + 3\frac{n}{t}(3\xi - 2\eta)\right)^{\beta-1}$$

$$\times \left((1+2\beta)\left[-\frac{B}{\gamma} + \frac{C_0}{\gamma}t^{3n\gamma(1+\alpha)}e^{\gamma(1+\alpha)}\right]^{\frac{1}{1+\alpha}} + (3+2\beta)\left(\frac{n}{t}(3\xi - 2\eta) + \frac{B}{\left[-\frac{B}{\gamma} + \frac{C_0}{\gamma}t^{3n\gamma(1+\alpha)}e^{\gamma(1+\alpha)}\right]^{\frac{\alpha}{1+\alpha}}}\right)\right), \tag{32}$$

$$p_{DE} = -\left(\left[-\frac{B}{\gamma} + \frac{C_0}{\gamma}t^{3n\gamma(1+\alpha)}e^{\gamma(1+\alpha)}\right]^{\frac{1}{1+\alpha}} + \frac{3B}{\left[-\frac{B}{\gamma} + \frac{C_0}{\gamma}t^{3n\gamma(1+\alpha)}e^{\gamma(1+\alpha)}\right]^{\frac{\alpha}{1+\alpha}}} + 3\frac{n}{t}(3\xi - 2\eta)\right)^{\beta} - \frac{n}{t}(3\xi - 2\eta). \tag{33}$$

We present the numerical analysis of evolution of the energy density and pressure of dark energy for the interacting model $Q = 3Hb\rho$, respectively, in figure 1 and 2.

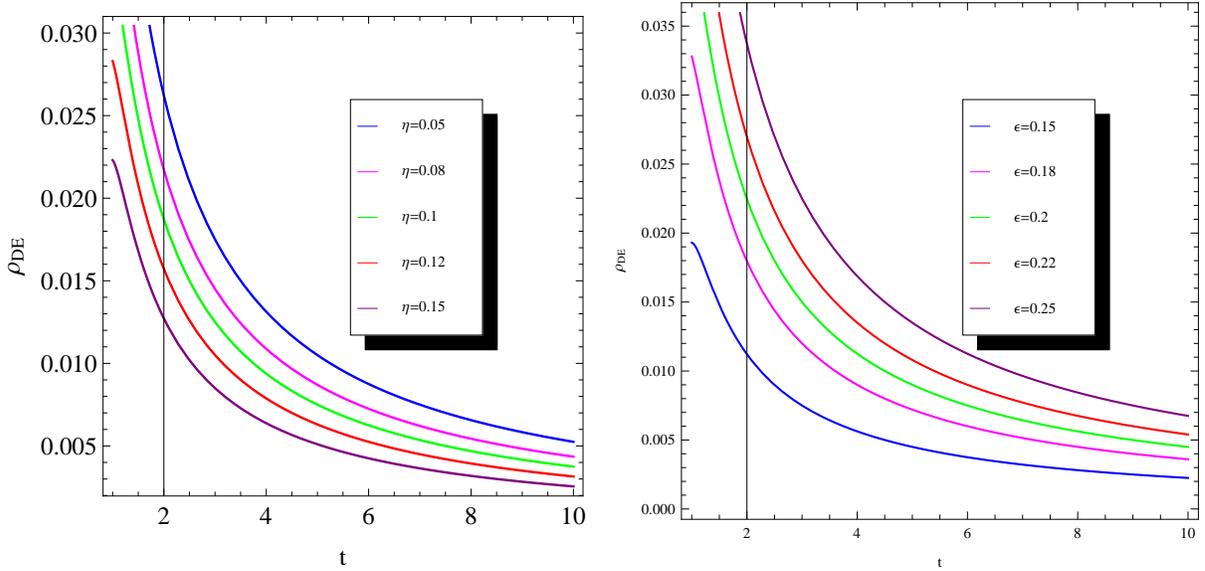

Figure 1: Plot of dark energy density versus cosmic time for $B = 3.4, \alpha = 0.5, \gamma = 2, C_0 = 1, \beta = 0.3, n = 0.5$. (First panel) We fix $\xi = 0.15$ and vary $\eta = 0.05, \eta = 0.08, \eta = 0.1, \eta = 0.12, \eta = 0.15$. (Second panel) We fix $\eta = 0.15$ and vary $\xi = 0.15, \xi = 0.18, \xi = 0.2, \xi = 0.22, \xi = 0.25$.

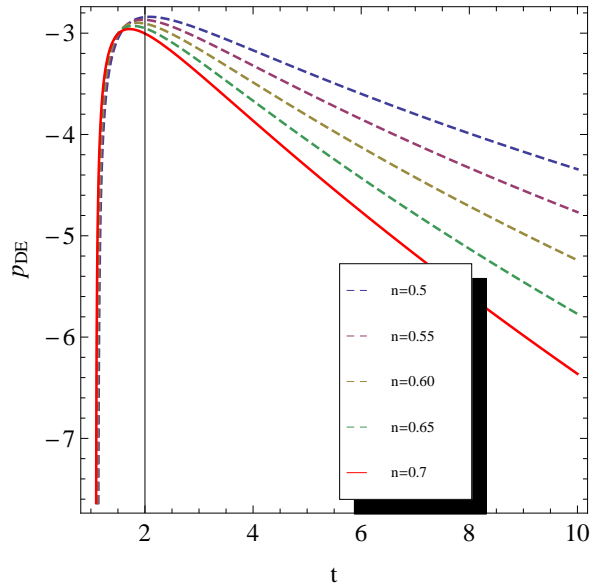

Figure 2: Plot of dark energy pressure versus cosmic time for
$B = 3.4, \alpha = 0.5, \gamma = 0.3, C_0 = 1, \beta = 0.3, \xi = 0.15, \eta = 0.05$.

### B. Interacting model $Q = q(\mu\dot{\rho} + 3bH\rho)$

We study in this section another type of interacting model. Hence, by using the continuity equation of conservation (18), one gets the following expression

$$\dot{\rho}\left(1 + \mu + \mu\frac{\dot{H}}{H^2}\right) + 3H\rho\left(1 + b + b\frac{\dot{H}}{H^2}\right) - \frac{3BH}{\rho^\alpha} = 0. \tag{34}$$

Within the special case of the scale factor (31), we can reformulate the last expression of the energy density of generalized Chaplygin gas as

$$\dot{\rho}\left(1 + \mu - n\mu\right) + 3\frac{n}{t}\rho\left(1 + b - nb\right) - 3B\frac{n}{t}\frac{1}{\rho^\alpha} = 0. \tag{35}$$

After integrating, the general solution of this equation yields

$$\rho = \left[-\frac{b_2}{b_1} + \frac{C'_0}{b_1}e^{b_1(1+\alpha)}t^{-\frac{b_1}{a_1}(1+\alpha)}\right]^{\frac{1}{1+\alpha}}, \tag{36}$$

where $a_1 = 1 + \mu(1 - n)$, $b_1 = 3n(1 + b - nb)$, $b_2 = -3nB$ and $C'_0$ the constant of integration.
From the equation of state (14), we deduct for this interacting model type, the pressure of the generalized Chaplygin gas as following form

$$p = -\frac{B}{\left[-\frac{b_2}{b_1} + \frac{C'_0}{b_1}e^{b_1(1+\alpha)}t^{-\frac{b_1}{a_1}(1+\alpha)}\right]^{\frac{\alpha}{1+\alpha}}}. \tag{37}$$

Using the equations (36) and (37), the energy and pressure of dark energy in $f(R,T)$ formalism yielding

$$\rho_{DE} = \left( \left[ -\frac{b_2}{b_1} + \frac{C'_0}{b_1} t^{-\frac{b_1}{a_1}(1+\alpha)} e^{b_1(1+\alpha)} \right]^{\frac{1}{1+\alpha}} + \frac{3B}{\left[ -\frac{b_2}{b_1} + \frac{C'_0}{b_1} t^{-\frac{b_1}{a_1}(1+\alpha)} e^{b_1(1+\alpha)} \right]^{\frac{\alpha}{1+\alpha}}} + 3\frac{n}{t}(3\xi - 2\eta) \right)^{\beta - 1}$$

$$\times \left( (1+2\beta) \left[ -\frac{b_2}{b_1} + \frac{C'_0}{b_1} t^{-\frac{b_1}{a_1}(1+\alpha)} e^{b_1(1+\alpha)} \right]^{\frac{1}{1+\alpha}} + (3+2\beta) \left( \frac{n}{t}(3\xi - 2\eta) + \frac{B}{\left[ -\frac{b_2}{b_1} + \frac{C'_0}{b_1} t^{-\frac{b_1}{a_1}(1+\alpha)} e^{b_1(1+\alpha)} \right]^{\frac{\alpha}{1+\alpha}}} \right) \right), \quad (38)$$

$$p_{DE} = -\left( \left[ -\frac{b_2}{b_1} + \frac{C'_0}{b_1} t^{-\frac{b_1}{a_1}(1+\alpha)} e^{b_1(1+\alpha)} \right]^{\frac{1}{1+\alpha}} + \frac{3B}{\left[ -\frac{b_2}{b_1} + \frac{C'_0}{b_1} t^{-\frac{b_1}{a_1}(1+\alpha)} e^{\gamma(1+\alpha)} \right]^{\frac{\alpha}{1+\alpha}}} + 3\frac{n}{t}(3\xi - 2\eta) \right)^{\beta} - \frac{n}{t}(3\xi - 2\eta) \quad (39)$$

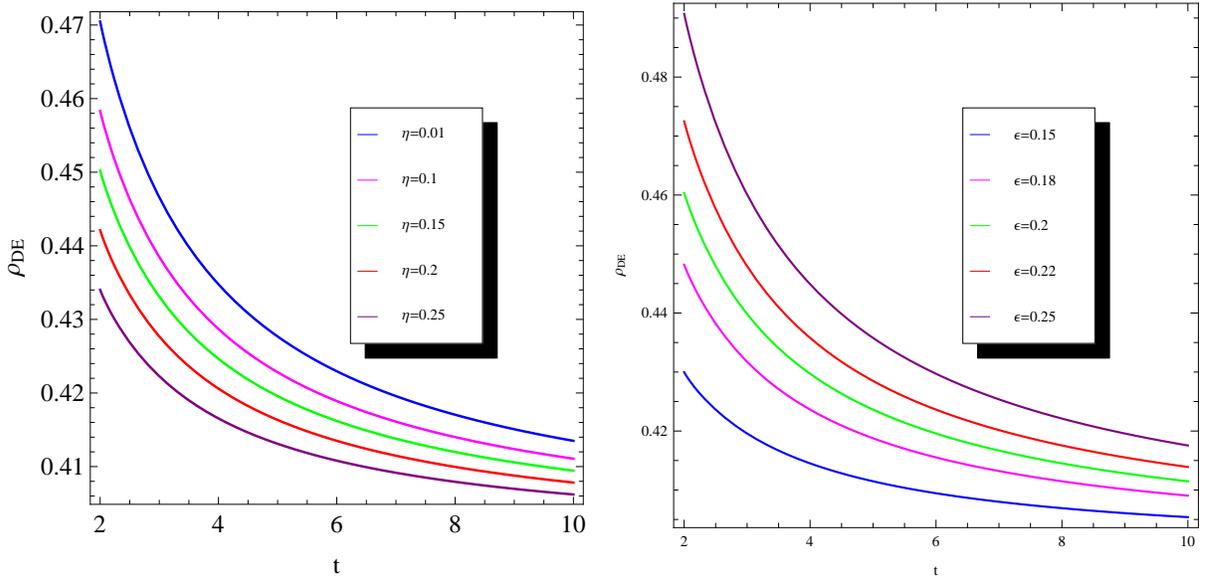

Figure 3: Plot of dark energy density versus cosmic time for $B = 3.4, \alpha = 0.5, \gamma = 2, C'_0 = 1, \beta = 0.3, n = 1.35, \mu = 1.5$. (First panel) We fix $\xi = 0.15$ and vary $\eta = 0.01, \eta = 0.1, \eta = 0.15, \eta = 0.2, \eta = 0.25$. (Second panel) We fix $\eta = 0.15$ and vary $\xi = 0.15, \xi = 0.18, \xi = 0.2, \xi = 0.22, \xi = 0.25$.

The figures 3 and 4 shows the evolution of energy density and pressure of dark energy for the interacting model $Q = q(\mu\dot{\rho} + 3bH\rho)$.

### C. Interacting model $Q = 3Hb\rho_{total}$

In this fact, we obtain through the equations (12), (14), (28) and (31) the differential equation in term of energy density of the generalized Chaplygin gas which can be expressed as

$$\dot{\rho} + 3\rho\frac{n}{t}(1 - b - 3b\beta) - 3B\frac{n}{t}\frac{1}{\rho^\alpha}(1 + b\beta) + 3b\beta\frac{n^2}{t^2}(2\eta - 3\xi) = 0. \quad (40)$$

Solving differential equation (40), we can obtained the time-dependent density. But this equation is difficult to solve analytically. We make the numerical analysis in order to find the energy density of generalized Chaplygin gas. Once the numerical analysis of the energy density of generalized Chaplygin gas be done, we can study the evolution of the energy density and pressure of dark energy given by equations (29) and (30) in figure 5.

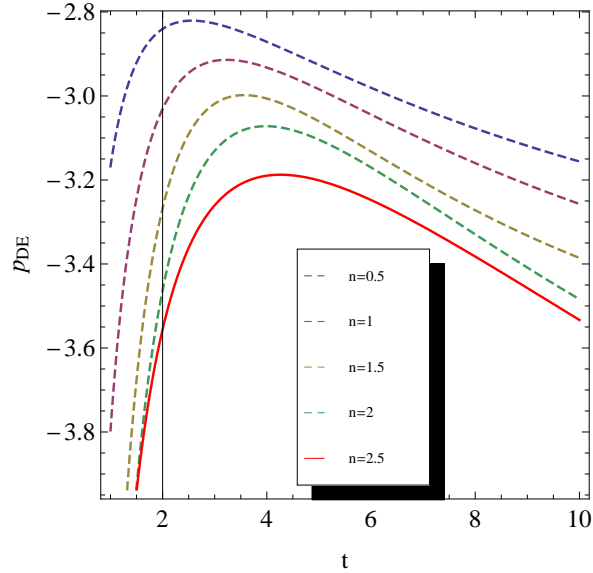

Figure 4: Plot of dark energy pressure versus cosmic time for
$B = 3.4, \alpha = 0.5, \gamma = 0.3, C_0 = 1.5, \beta = 0.3, \mu = 1.5, \xi = 0.15, \eta = 0.05$

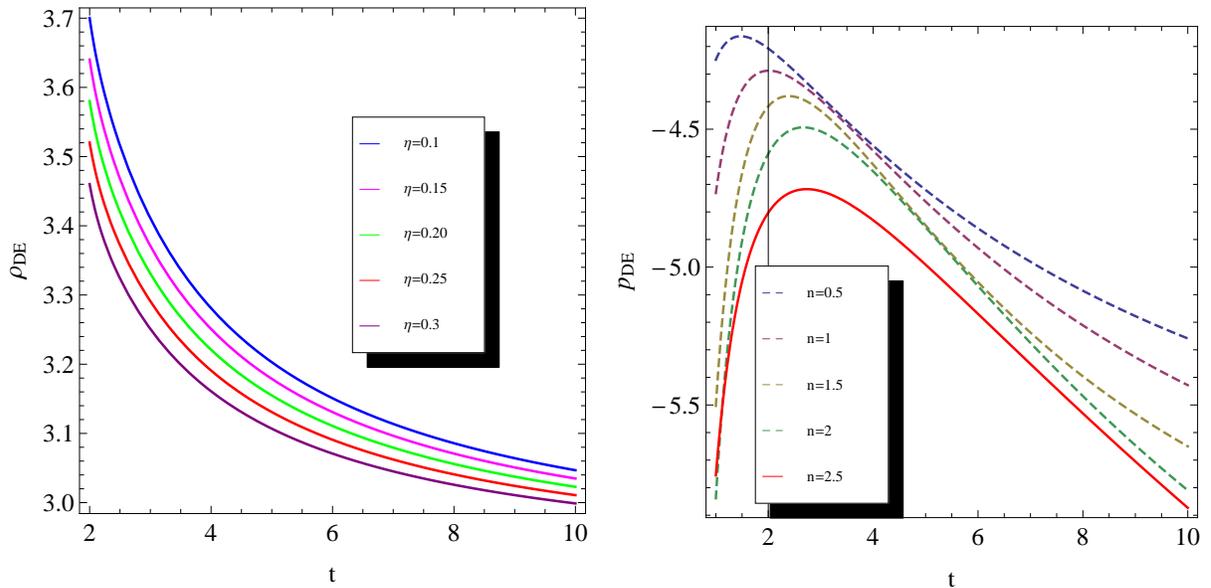

Figure 5: Plot of dark energy density and pressure for $B = 3.4, \alpha = 0.5, C'_0 = 1.5, \beta = 1, b = 1.5, \mu = 2.5$. (First panel) We fix $n = 1.2, \xi = 0.5$ and vary $\eta = 0.1, \eta = 0.15, \eta = 0.2, \eta = 0.25, \eta = 0.3$. ( Second panel) We fix $\xi = 0.05$, $\eta = 0.05$ and vary $n = 0.5, n = 1, n = 1.5, n = 2, n = 2.5$.

Dependent on three type of interacting models $Q$ considered, we note that free parameters play the role of an important to plot the cosmological parameters such as energy density and pressure of dark energy according with observational data, in which the motivation is based on positivity energy density and negativity pressure of dark energy. We remark that when the cosmic time evolve, the dark energy density increases with bulk viscosity while shear viscosity decreases value of dark energy density. We observe also that the accelerating expansion of the universe filled with viscous generalized Chaplygin gas decays the pressure of dark energy for the power-law scale factor.

In general case of the $f(R,T)$ formalism, the first generalized Friedmann equation (12) yields

$$3H^2 = \kappa_{eff}^2 \left( \rho + \frac{1}{1+f_T} \left[ \frac{1}{2}\big(f(R,T) - Rf_R\big) - 3\dot{R}H f_{RR} + \bar{p} f_T \right] \right), \tag{41}$$

where $\kappa_{eff}^2 = \frac{f_R}{1+f_T}$ and $\rho_{DE} = \frac{1}{1+f_T}\left[\frac{1}{2}\big(f(R,T) - Rf_R\big) - 3\dot{R}H f_{RR} + \bar{p} f_T\right]$.

Assuming without special cases of scale factor (31), we obtain after using the equations (9), (14) and (41) the general form of the differential equation (40) for $f(R,T)$ model as

$$\dot{\rho}(1+f_T) + 3H\rho(1-b)(1+f_T) - 3BH\frac{1}{\rho^\alpha}(1+f_T - bf_T) - \frac{3Hb}{2}f(R,T) \tag{42}$$

$$-9bH(2H^2 + \dot{H})f_R - 54bH^2(4\dot{H}H + \ddot{H})f_{RR} - 3bH^2(2\eta - 3\xi)f_T = 0. \tag{43}$$

## V. CONCLUSION

In this paper, we studied generalized Chaplygin Gas interacting with $f(R,T)$ gravity in presence of shear and bulk viscosities. By employing that the contents of universe is dominated by dark sector i.e the generalized Chaplygin gas (dark matter) and dark energy density, we obtained through the modified Friedmann equations the time-dependent energy density and pressure of generalized Chaplygin gas and dark energy, respectively for special model $f(R,T) = R + 2T^\beta$. We obtained effects of shear and bulk viscosities on dark energy density for three interacting model $Q$, which are decreasing and increasing, respectively. We showed that the pressure of dark energy decays with accelerating expansion of universe and we concluded that the viscous parameters help to obtain solution coincident with observational data.

---